\documentclass[twocolumn]{aastex63}

\shorttitle{Skynet's New Observing Mode: The Campaign Manager}
\shortauthors{Dutton et al.}
\graphicspath{{./}{figures/}}

\begin{document}

\title{Skynet's New Observing Mode: The Campaign Manager}

\author[0000-0003-3144-7369]{Dylan A. Dutton}
\author[0000-0002-8559-5888]{Daniel E. Reichart}
\author{Joshua B. Haislip}
\author[0000-0003-3642-5484]{Vladimir V. Kouprianov}
\author[0000-0002-7683-7267]{Omar H. Shaban}
\author[0000-0002-1906-1167]{Alan Vasquez Soto}

\affiliation{Department of Physics and Astronomy \\
University of North Carolina at Chapel Hill \\
Chapel Hill, NC 27599-3255, USA}

\begin{abstract}

Built in 2004, the Skynet robotic telescope network originally consisted of six 0.4-m telescopes located at the Cerro-Tololo Inter-American Observatory in the Chilean Andes. The network was designed to carry out simultaneous multi-wavelength observations of gamma-ray bursts (GRBs) when they are only tens of seconds old. To date, the network has been expanded to $\approx$20 telescopes, including a 20-m radio telescope, that span four continents and five countries. The Campaign Manager (CM) is a new observing mode that has been developed for Skynet. Available to all Skynet observers, the CM semi-autonomously and indefinitely scales and schedules exposures on the observer's behalf while allowing for modification to scaling parameters in real time. The CM is useful for follow up to various transient phenomena including gravitational-wave events, GRB localizations, young supernovae, and eventually, sufficiently bright Argus Optical Array and Large Synoptic Survey Telescope events.
\\

\end{abstract}

%% -- new sections -- %%
\section{Introduction} \label{sec:intro}
Due to the unpredictable nature for both the location and
time of transient events, using a single telescope, or even
multiple telescopes at a single site, leaves the execution of an
observation entirely up to happenstance. Even if the event
occurs in the same hemisphere as the telescope, continuous
observing will not be possible using only one telescope site.
This may result in missing critical information, particularly for
fast fading events such as gamma-ray burst (GRB) afterglows
or kilonovae. To avoid this, a well-distributed global network
of telescopes is needed to ensure that continuous monitoring of
transients can occur. With near-continuous coverage of the
southern hemisphere as well as built in coverage redundancy in
North America, South America, and Australia, the Skynet
robotic telescope network is ideal for such events.

Skynet is a sophisticated telescope control and queue
scheduling software that simultaneously controls a global network
of telescopes, allowing them to function individually or as an
integrated whole. Furthermore, Skynet can control most commercially available telescope hardware, and provides participating
institutions with easy-to-use web and API interfaces. Participating
institutions are not charged, but instead contribute 10\% of each of
their telescopes’ time for Director Discretionary science and
education. Telescopes are regularly being added to the Skynet
Robotic Telescope Network, which has now grown to number
$\approx$20 telescopes, ranging in size from 14" to 40", with another $\approx$10
scheduled to join in the near future.

The Panchromatic Robotic Optical and Monitoring Polarimetry Telescopes (PROMPT) are a subset of the Skynet
Robotic Telescope Network, consisting of the network’s
highest-quality telescopes at its highest-quality sites. Originally
only at Cerro-Tololo Inter-American Observatory (CTIO),
PROMPT now spans five dark sites, in Chile, Australia (for
near-continuous observing in the southern hemisphere), and
Canada (for full-sky coverage). Many of these are original, and
identical, telescopes that have been redeployed from CTIO.

For nearly two decades, Skynet has proven successful at
observing a wide variety of transients. Skynet/PROMPT was
originally built to carry out simultaneous, multi-wavelength
observations of GRBs when they are only tens of seconds old (to
date, Skynet has observed 88 GRBs within 15-70s (90\% range)
of spacecraft notification, detecting 50 optical afterglows on this
timescale). In addition to being used to study GRBs (Reichart
et al. 2005; Haislip et al. 2006; Dai et al. 2007; Updike et al.
2008; Nysewander et al. 2009; Cano et al. 2011; Cenko et al.
2011; Bufano et al. 2012; Jin et al. 2013; Martin-Carrillo et al.
2014; Morgan et al. 2014; Friis et al. 2015; Bardho et al. 2016;
De Pasquale et al. 2016; Melandri et al. 2017), Skynet is being
used to study Gravitational Wave (GW) sources, (Abbott et al.
2017a, 2017b; Valenti et al. 2017; Yang et al. 2017, 2019; Abbott
et al. 2020), fast radio bursts, blazars, supernovae (SNe),
supernova remnants, novae, pulsating white dwarfs and hot
subdwarfs, a wide variety of variable stars, a wide variety of
binary stars, exoplanetary systems, trans-Neptunian objects and
Centaurs, asteroids, and near-Earth objects. Skynet data are now
published in peer-reviewed journals every $\approx$20 days.

The Campaign Manager (CM) is a new observing mode that
has been developed to improve Skynet's capabilities for
observing transients such as GW events, GRB localizations,
young SNe, and, once construction is completed, sufficiently
bright Argus Optical Array (Law et al. 2021) and Large
Synoptic Survey Telescope events. This new semi-autonomous
observing mode designed for transients allows for auto-scheduling 
and auto-scaling of exposures on all Skynet
telescopes. Additionally, the CM allows for real-time updates
to the observing parameters so that exposure lengths can be
appropriately updated as new information is learned about the
target. This is a significant improvement over the existing
observing method that requires observations to be canceled and
rescheduled using the new parameters. The CM is available to
any Skynet observer, but is most effective when coupled with
target-of-opportunity

\begin{figure*}
    \includegraphics[scale=0.58]{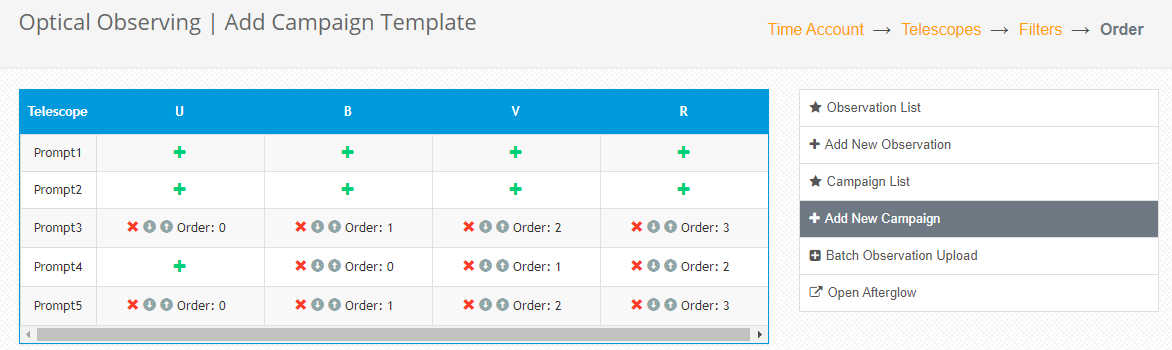}
    \caption{\textit{The displayed list of selected telescopes and filters for a hardware template. The ''green plus''
    indicates that the given filter (column) is not installed on the given telescope (row). If clicked, the CM will
    schedule the particular combination once/if the filter is installed.}}
\end{figure*}

An overview of Skynet's major components is given in
Section 2. A description of the CM’s front-end web-interface is
given in Section 3. A review of the back-end structure of the
CM is given in Section 4. A scientific application of the CM is
given in Section 5 section via the Skynet research group’s
current use of the CM.

%% -- new section -- %%
\section{Skynet's Software Structure} \label{sec:skynet}
The major components of Skynet include: (1) the Optical
Server, (2) the Terminator client, and (3) the website. The
entirety of the Skynet system, with the exception of the
Terminator which is installed on each telescope's control
computer, runs on servers that are located at and maintained by
the University of North Carolina at Chapel Hill.

The Optical Server is responsible for monitoring the states of
the telescopes as well as dispatching observations. To do this,
the Optical Server sends commands to the telescope client. In
turn, the Terminator monitors and handles the management of
each telescope’s equipment including the mount, camera, filter
wheel, dome, and weather monitoring system. The Terminator
is capable of detecting, and in some cases resolving, hardware
issues. If the Terminator is unable to resolve an issue, an alert is
sent to the Skynet team

Skynet currently uses Web2Py which is a popular ModelView-Controller (MVC) 
architectural pattern for its web
development framework. Access to the website (\url{https://skynet.unc.edu}) 
requires a Skynet account. Currently, accounts
are not available to the general public, but are limited to
students, researchers, and those with ownership shares on
telescopes. Observers use the website to submit observations,
check the status of telescopes, and retrieve their images.

%% -- new section -- %%
\section{Campaign Manager User Interface} \label{sec:interface}
Skynet observers have access to the campaign manager via
both the Skynet web-interface and the Skynet API. All Skynet
observers are provided an API key that can be found under
their account settings. Additionally, ready-to-use Python scripts
that interact with the API are available upon request. For
observers that prefer a user interface, the CM can be used via
the Skynet website. The process of using the CM involves three
main the steps: (1) defining the hardware template, (2) adding
an observation, and (3) updating the observing parameters as
desired.

%% Hardware Templates
\subsection{The Hardware Template} \label{sec:hardware}
The hardware template is a pre-defined, ordered list of the
desired telescopes and filters that are referenced by the CM
during an observation. In the event of a rapidly fading transient,
it is desirable to begin taking exposures immediately after a
candidate for the event has been discovered. For this reason,
the hardware template can be created at any time prior to an
active event in order to shorten the time between the start of an
event and the start of an observation.

The observer can add a hardware template by navigating to
the “Add Campaign Template” after clicking on the “Optical
Observing” under the “My Observatory” tab in the main
navigation pane. The observer's first step in creating a template
is selecting the desired filters for the observation. Available
filters include Sloan and Johnson/Cousins as well as narrowband, astrophotography,
and special purpose filters.

Following filter selection, the observer is prompted to select the
desired telescopes. In addition to the list of available telescopes,
the observer is presented with an option to automatically add new
telescopes as they become available. If this option is selected, the
CM will add any new telescopes, along with the filters that were
selected in the previous step, that the user has been granted access
to that are not in the hardware template. As an example, if the
observer only had access to PROMPT-1 and PROMPT-2 at the
time of creating the hardware template, but was granted access to
PROMPT-3 at some later time, then the CM would add
PROMPT-3 to the template and schedule exposures on all three
telescopes at the time of the observation.

After selecting the desired telescopes and filters, the observer
is directed to a new web page containing a table that presents
the telescope and filter combinations (see Figure 1). If a
particular telescope/filter combination does not exist, i.e., the
filter is not installed on the given telescope, then the table entry
is populated with a green plus sign. This gives the observer the
option to add the combination to the template even if it does not
exist. If the filter is installed on the telescope at the time of the
observation, then the CM will schedule exposures using the
combination. If the filter is still not installed, then the CM will
simply ignore the combination when scheduling exposures.This option 
removes the need for the observer to continuously
monitor which filters are installed on the Skynet telescopes.

In the same table, the observer has the option to specify the order
that the CM schedules the filtered exposures for each telescope.
Although Skynet is a heterogeneous network of robotic telescopes,
there are subsets, e.g., PROMPT, that have identical equipment.
Allowing the observer to specify the order prevents identical
telescopes from taking identical exposures at the same time. This
prevents the collection of redundant information and maximizes the
efficiency of using the Skynet network of telescopes.

%% Adding an Observation
\subsection{Adding an Observation} \label{sec:observation}
For Skynet to begin scheduling and dispatching exposures,
an observation must be submitted. The observer can add an
observation by navigating to the “Add New Observation” after
clicking on the “Optical Observing” under the “My Observatory” tab in the main navigation pane.

When selecting the targets coordinates, the observer has the
option to use the target lookup feature that uses local catalogs
and SIMBAD (Wenger et al. 2000) to auto-populate the right
ascension (R.A.) and declination (decl.) of the target. However,
since the CM is designed to be used for transient targets, it is
unlikely that the target will be cataloged. As an alternative, the
observer can manually enter the sky coordinates.

In addition to the coordinates, event-specific parameters are
required to auto-scale exposure lengths. These include: (1) a
desired signal-to-noise ratio (S/N) per exposure; (2) an assumed
brightness in an observer-selected filter at an observer-selected
time, post-event; (3) an assumed temporal index (both power-law
and exponential options are available); (4) an assumed spectral
index; and (5) a maximum acceptable exposure duration.\footnote{If an 
exposure would exceed a telescope's recommended maximum tracking duration,
a block of exposures is scheduled instead} 

The TOO option must also be specified before submitting the
observation. Though the CM is available without TOO, it is
most effective when paired with it. Enabling this option gives
the observer exclusive access to the specified telescopes for the
duration of the observation. If multiple TOO observations from
different observers are active on the same telescopes, Skynet’s
load balancer is capable of handling such scenarios.

Finally, a configurable option to delay filter sequences is also
available before submitting the observation. Adding a delay
between filter sequences is useful for transients that vary slowly
with time. The observer may choose from a fixed delay length
or a delay length that scales based on the time since the event.
With these parameters defined, the observation can be
submitted and exposures will be immediately dispatched to
the selected telescopes using the selected filters.

\begin{figure}
    \includegraphics[scale=0.4]{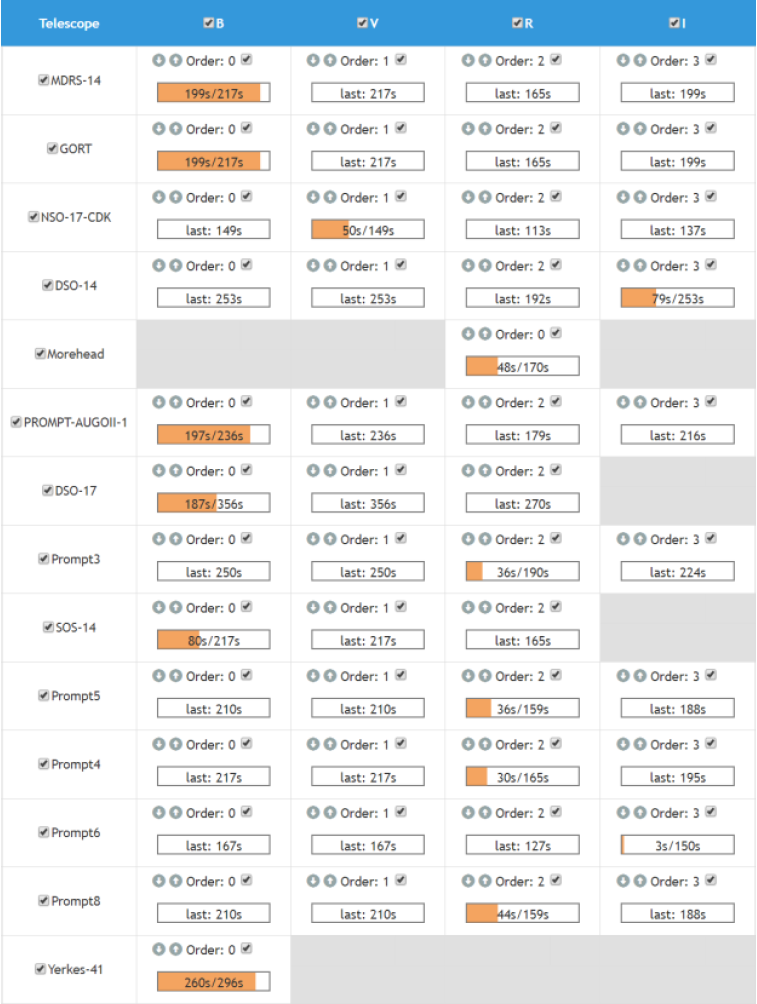}
    \caption{\textit{The ``Live Update'' display for a large-scale campaign using Johnson/Cousins filters. Progress bars 
    with orange indicate an ongoing exposure. Radio buttons with check marks indicate that the particular telescope/filter 
    combination has not been paused by the observer. Unchecking the radio button will cause the CM to cease scheduling exposures 
    for that particular combination until rechecked.}}
\end{figure}

%% Monitoring an Observation
\subsubsection{Monitoring the Observation} \label{sec:monitor}
After the observation has been submitted, the CM automatically runs through 
all filter lists on all telescopes in the
hardware template, automatically scaling exposure or exposure-block
durations from scope to scope, from filter to filter,
and as a function of time, given each telescope/filter
combination's typical limiting magnitude and the assumed
temporal and spectral indices, to achieve similar S/N per
exposure or exposure block. 

To monitor the exposures being scheduled, a “Live Update”
web-page has been developed specifically for use with the CM.
A link to the “Live Update” page is accessed by clicking the
link under the observation info pane on the observation page.
The “Live Update” page (see Figure 2) displays a table similar
to that of the hardware page. However, the table cells now
display, along with the order, a progress bar that shows not
only the exposure length, but the current progression of that
exposure for each telescope/filter combination. Each table cell
also allows the observer to modify the order of the filter sequence in 
real time, if desired. Monitoring the observation
can also be done via the Skynet API, if desired.

In addition to the telescope/filter table, there is a web-form
positioned above the table that displays all of the event-specific
parameters as well as their corresponding values. As new
information is learned about the target, such as its brightness at
a certain time in a certain filter, or its temporal and spectral
indices, these parameters can be modified and the CM will
immediately update upcoming exposure lengths and display the
changes in the table.

Finally, the observer can discontinue use of particular
telescopes, particular filters, or combinations thereof (or add
them back in) in real time. Otherwise, each telescope/filter
combination automatically discontinues once the maximum
acceptable exposure or exposure-block duration is exceeded.

\begin{figure*}
    \includegraphics[scale=0.55]{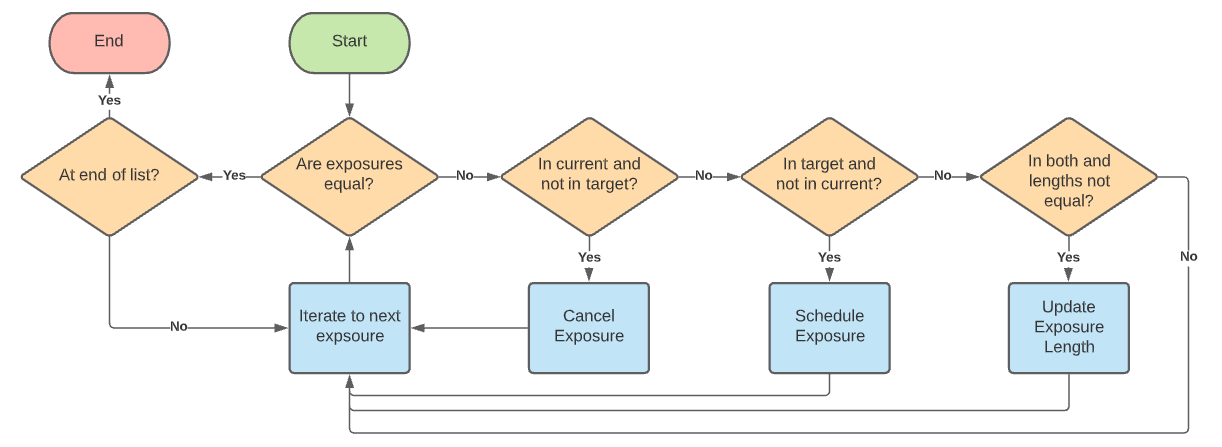}
    \caption{\textit{The matching algorithm comprising of the CM's third and forth main objectives as described in $\S$4.1.
    Following the generation of the current list, i.e., the list of currently scheduled exposures, and the target list,
    i.e., the list of exposures that should be scheduled based on any recent user-inputted changes, the CM iterates through 
    both lists, compares their properties, and modifies the current list to match the target list.}}
\end{figure*}

%% -- new section -- %%
\section{The Campaign Manager} \label{sec:cm}
The CM is built into the existing Skynet source code and is
almost exclusively contained within the MVC structure. That
is, the data-related logic is handled in the model, the data
manipulation and processing are handled in the controller, and
the user interface (UI) logic is handled in the views. The only
CM code not contained in the MVC is the calling routine. The
calling routine is contained in a script named the “client
monitor.” An instance of the client monitor is created for each
live telescope connected to Skynet and is responsible for
handling the telescope's exposure queue and sending the
appropriate commands to the Terminator. Hence, the CM is
called on a per-telescope basis and is only responsible for
handling exposures from the telescope associated with the
given client monitor instance.

%% Campaign Manager Goals
\subsection{Objectives} \label{sec:cmobj}
The CM algorithm has four main objectives: (1) generate a
list of exposures that should be scheduled based on the
observer's selected telescopes/filters, (2) calculate the exposure
length for each exposure using the observer's event-specific
parameters, (3) compare the list of currently scheduled
exposures to the list of exposures in (1), and (4) add/cancel/update 
exposures as appropriate according to (3). (see Figure 3)
for a visualization of objectives (3) and (4).

%% Activating Criteria 
\subsection{Activation} \label{sec:act}
The CM does not start up and run indefinitely once a
telescope connects to Skynet. Instead, the main loop in the
client monitor, which runs until the telescope is disconnected,
continuously checks if there are any active observations and if
the observer has specified that the observation should be
handled by the CM. The observer can flag an observation for
the CM via either the Skynet web interface or the API when
adding the observation.

Once activated, the CM schedules four minutes of exposures
based on the observers inputted parameters. Since the CM does
not run indefinitely as a standalone script, it must be repeatedly
called in order to update the existing exposure (if there is an
update to the parameters by the observer) or to schedule new
exposures beyond the initial four minute schedule. By default,
the CM is called every 60s to perform the four objectives listed
in Section 4.1. Additionally, the CM is called immediately after
an observer updates any of the input observation parameters.
The 60s timer is reset after an update to the parameters is
detected.

%% Exposure Lists
\subsection{Exposure Lists} \label{sec:explists}
To semi-autonomously schedule exposures, the CM's first
task is to create a list of exposures corresponding to the ordered
filters that the observer specified in their hardware template. If
the CM continuously ran as a background script, then this
would be as simple as checking a list that gets appended each
time an exposure is scheduled. However, since the CM is called
periodically, it must first determine which exposures have
already been scheduled and are waiting to be executed such
that the observer's specified order can be preserved.

The process of determining which exposures have already
been scheduled is further complicated since the CM is
repeatedly scheduling the same sequence of filters. When
calculating the exposure lengths (Section 4.4) it is crucial to
know how many times the filtered exposure has been scheduled
due to the time dependency of the scaling equation. To remove
this degeneracy, a unique identifier called the “iteration ID” is
defined. The iteration ID is incremented each time that a full
sequence of exposures is scheduled. For example, if the
observer requested the filters \textit{B}, \textit{V}, and\textit{R}, 
then each exposure in that sequence would have the same iteration ID. However, the
iteration ID would be incremented the next time they are
scheduled.

The CM will schedule ordered exposures until at least two
exposures and at least four minutes are scheduled. Though the
time limit of four minutes is semi-arbitrary, both conditions are
necessary in ensuring that the observation is executed without
interruption (see Figure 4). The client monitor is constantly
checking for exposures that need to be executed on a given
telescope. If the CM were to schedule a single exposure with an
exposure length of 15 s, then there would be a 45 s gap in
which there are no exposures scheduled for the observation. To
prevent an idle telescope, the client monitor would begin
executing exposures for another observation even if the
observer had specified TOO. The telescope would then slew
to the new target only to have slew back once the CM
scheduled another exposure. Requiring at least two exposures
and a minimum time buffer of at least 60s prevents this.

\begin{figure}[h]
    \hspace*{0.9cm}
    \includegraphics[scale=0.6]{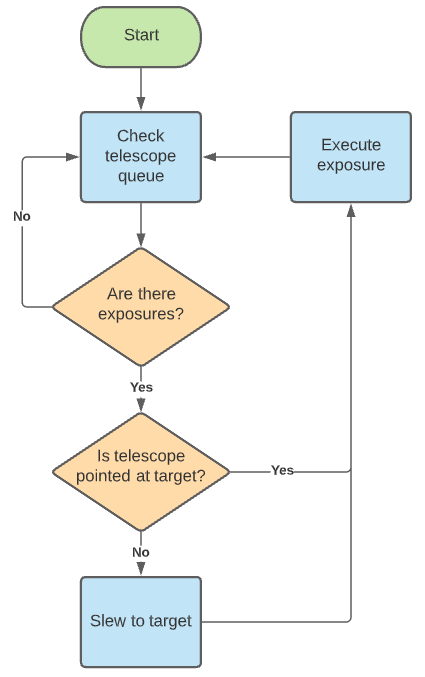}
    \caption{\textit{A greatly simplified client monitor algorithm emphasizing the importance of the CM scheduling
    exposures into the future. If the CM did not schedule at least two exposures with a minimum 60-second time buffer,
    then the telescope would repeatedly slew to other targets wasting valuable observing time.}}
\end{figure}

%% Scaling Equation
\subsection{The Scaling Equation} \label{sec:eq}
Skynet employs a credit-based currency system that
observers use to schedule their observations. However, since
Skynet is a heterogenous network, one credit cannot be easily
mapped to a fixed unit of observing time. For example, to
achieve a S/N for a given target, a 1.0 m telescope in CTIO
would need to expose for a shorter duration than the 40 yr old,
24 inch Morehead telescope located in downtown Chapel Hill.
For this reason, mapping one credit to one observing second, or
any other fixed unit of time, does not make sense.

Instead, Skynet charges one credit for a one-second exposure
on Skynet's PROMPT-5 telescope and scales based on the
observer’s selected telescope's efficiency relative to PROMPT5. 
To achieve this, Skynet uses a cron job to calculate the
efficiencies of each telescope once per week using the most
recent, open filtered exposures. These efficiencies, $\varepsilon_{t}$, are
defined as the inverse of the exposure time, $t_{exp}$, needed to
achieve a limiting magnitude, $m$, of 20 at a $S/N$ of 5:

\begin{equation} \label{eq:eff}
    \varepsilon_{t} = \frac{\left[ (S/N) / 5 \right]^{2} 
        \cdot log_{10}[(m - 20) / 2.5]}{t_{exp}}
\end{equation}

Here, the assumption is made that the source of interest is
sufficiently bright enough that the noise is purely Poisson.
Thus, the $S/N$ factor in Equation (1) must be squared. Outlier
rejection is performed using Robust Chauvenet Rejection
(Maples et al. 2018) to eliminate anomalous data points due to
weather, hardware issues, or any other source of contamination.
Equation (1) is used as the starting point for scaling exposure
lengths.

However, since the CM is interested in calculating exposure
lengths, not efficiencies, we start by taking the inverse of
Equation (1). Since telescope efficiencies are calculated using
open filters only, a factor must be added to account for the
efficiencies of the observer's selected filter. This accounts for
the remaining hardware dependencies, but if left as is, the
scaling equation still fails to account for event-specific
parameters. To account for these, a time-dependent factor, as
well as a frequency-dependent factor are added, yielding:

\[
    t_{exp} = \left[ (S/N) / 5 \right]^{2} 
        \cdot log_{10}[(m - 20) / 2.5] 
\]
\begin{equation} \label{eq:full}
        \times \left( \frac{\varepsilon_{t}}{\varepsilon_{f}} \right)
        \left( \frac{t_{exec} - t_{event}}{t_{ref}} \right)^{-\alpha}
        \left( \frac{\nu_{f}}{\nu_{f, ref}} \right)^{-\beta}
\end{equation}

The filter efficiency, $\varepsilon_{f}$, is calculated and stored as a value
between 0 and 1 in the Skynet database. On the other hand, the
telescope efficiency is stored as an inverse and can achieve a
value higher than 1 as a result. The time-dependent factor
normalizes the time between the execution of the exposure,
$t_{exec}$, and the start of the transient event, $t_{event}$, and then scales it
according to the assumed temporal index of the source, $\alpha$. The
frequency-dependent factor normalizes the center frequency of
selected filter, $\nu_{f}$, to the center frequency of the reference filter,
$\nu_{f,ref}$, and then scales it according to the assumed spectral index
of the source, $\beta$.

Similarly, an exponential scaling function that makes use of
the same parameters is also available

\[
    t_{exp} = \left[ (S/N) / 5 \right]^{2} 
        \cdot log_{10}[(m - 20) / 2.5] 
\]
\begin{equation} \label{eq:full}
        \times \left( \frac{\varepsilon_{t}}{\varepsilon_{f}} \right)
        \left( \exp\left[-\alpha\frac{t_{exec} - t_{event}}{t_{ref}} \right]\right)
        \left( \frac{\nu_{f}}{\nu_{f, ref}} \right)^{-\beta}
\end{equation}

\noindent The observer has the option to specify either a power or exponential function while adding the observation.

%% Stopping Conditions
\subsection{Stopping the Campaign Manager}
There are four independent stopping conditions for the CM
to cease scheduling exposures and cancel the observation. The
stopping conditions include: (1) an observer expends all of
their available credits, (2) all of the telescope/filter combinations have been masked, (3) the observer selected cancel-after
date has been reached, or (4) the observer manually cancels the
observation.

Masking available telescope/filter combinations can be done
manually by the observer, or automatically by the CM if
exposure lengths exceed the observer-inputted maximum
allowable length. In most cases, the observer will manually
end the observation once they have collected their data. The
cancel-after date is intended as a fail-safe, cut-off date to
prevent the observer from accidentally expending all of their
credits.

%% -- new section -- %%
\section{Discussions \& Conclusions} \label{sec:disc}
The recent emergence of multi-messenger astronomy has
resulted in multiple large-scale observing campaigns in an
attempt to better understand some of the most violent processes
in the universe. Following a successful discovery of the
electromagnetic counterpart to GW170817 (e.g., Abbott et al.
2017a), Skynet sought to improve its follow-up observing
capabilities. Thus, development began on the campaign
manager in 2018 with the goal expanding Skynet's existing
capabilities to observe transients to better image optical
counterparts to compact binary coalescences.

Before the CM can be utilized for follow-up observations,
candidates to such events must first be discovered. Currently,
Skynet has two internal methods, and one external method, for
candidate detection. If a candidate is detected via an internal
method, a detection alert will be distributed to the greater
astronomy community through the Gamma-ray Coordinates
Network and the CM will be immediately and automatically
triggered via the API. In the case of an external detection by
another research group, the CM will be triggered manually
using the candidate's sky coordinates.

The first internal discovery method utilizes Distance Less
Than Forty (DLT40), which is a young, less than 100 Mpc
(previously 40 Mpc) supernova search that uses Skynet's
identical PROMPT telescopes at our Chilean, Western
Australian, and Canadian sites to observe $\approx$400-600 galaxies
per night per site. DLT40's narrow-field, targeted search
approach was proven successful with their co-discovery of the
GW170817 optical counterpart (Valenti et al. 2017).

The other internal discovery method utilizes the Evryscopes
(Law et al. 2015), two large arrays of small telescopes, in Chile
and California, that cover the entire visible sky in a single
exposure, repeatedly imaging the sky every two minutes and
stacking images to achieve depth. The two Evryscopes,
designed, built, deployed and operated by Law's group at the
University of North Carolina at Chapel Hill, use massproduced compact CCD cameras, mass-produced compact
camera lenses, and a novel camera mounting scheme to make
robotic telescopes with a total of 1.3 gigapixels covering the
sky. The Evryscopes were not utilized in the 2017 kilonova
discovery. As such, their images of 16,000 square degrees in
each exposure may greatly reduce the time between the LIGO
detection and the activation of the CM if there is a sufficiently
bright, early counterpart.

Although designed with LIGO/Virgo/KAGRA events in
mind, the CM is well equipped to handle observations of
various other transient phenomena. Skynet will be using the
CM to observe GRBs between GW events

\section{Acknowledgments}
We gratefully acknowledge the support of the National Science Foundation, specifically through the Astronomy and Astrophysics Grant 2007853.

\section{References}
Abbott, B. P., Abbott, R., Abbott, T. D., et al. 2017a, ApJL, 848, L12\\
Abbott, B. P., Abbott, R., Abbott, T. D., et al. 2017b, Natur, 551, 85\\
Abbott, B. P., Abbott, R., Abbott, T. D., et al. 2020, Phys. Rev. D, 101, 8\\
Bardho, O., Gendre, B., Rossi, A., et al. 2016, MNRAS, 459, 508\\
Bufano, F., Pian, E., Sollerman, J., et al. 2012, ApJ, 753, 67\\
Cano, Z., Bersier, D., Guidorzi, C., et al. 2011, MNRAS, 413, 669\\
Cenko, S. B., Frail, D. A., Harrison, F. A., et al. 2011, ApJ, 732, 29\\
Dai, X., Halpern, J. P., Morgan, N. D., et al. 2007, ApJ, 658, 509\\
De Pasquale, M., Oates, S. R., Racusin, J. L., et al. 2016, MNRAS, 455, 1027\\
Friis, M., De Cia, A., Krühler, T., et al. 2015, MNRAS, 451, 167\\
Haislip, J. B., Nysewander, M. C., Reichart, D. E., et al. 2006, Natur, 440, 181\\
Jin, Z.-P., Covino, S., Della Valle, M., et al. 2013, ApJ, 774, 114\\
Law, N., Corbett, H., Vasquez Soto, A., et al. 2021, AAS Meeting Abstracts,
53, 235.02\\
Law, N. M., Fors, O., Ratzloff, J., et al. 2015, PASP, 127, 234\\
Maples, M. P., Reichart, D. E., Konz, N. C., et al. 2018, ApJS, 238, 2\\
Martin-Carrillo, A., Hanlon, L., Topinka, M., et al. 2014, A\&A, 567, A84\\
Melandri, A., Covino, S., Zaninoni, E., et al. 2017, A\&A, 607, A29\\
Morgan, A. N., Perley, D. A., Cenko, S. B., et al. 2014, MNRAS, 440, 1810\\
Nysewander, M., Reichart, D. E., Crain, J. A., et al. 2009, ApJ, 693, 1417\\
Reichart, D., Nysewander, M., Moran, J., et al. 2005, NCimC, 28, 767\\
Updike, A. C., Haislip, J. B., Nysewander, M. C., et al. 2008, ApJ, 685, 361\\
Valenti, S., Sand, D. J., Sheng, Y., et al. 2017, ApJL, 848, L24\\
Wenger, M., Ochsenbein, F., Egret, D., et al. 2000, A\&AS, 143, 9\\
Yang, S., Sand, D. J., Valenti, S., et al. 2019, ApJ, 875, 59\\
Yang, S., Valenti, S., Cappellaro, E., et al. 2017, ApJL, 851, L48\\

%\bibliography{campaignmanager}{}
%\bibliographystyle{aasjournal}

\end{document}